\documentclass[prb,showpacs,twocolumn]{revtex4}
\usepackage{graphicx}
\usepackage{subfigure}
\begin{document}

\title{Possible superconducting symmetry on doped $J_1$-$J_2$ model}

\author{Huai-Xiang Huang, Ying Jiang, Guo-Hong Yang}
\affiliation { Department of Physics, Shanghai University,
ShangHai, 200444, China}

\date{\today}

\begin{abstract}
By making use of renormalized mean-field theory, we investigate
possible superconducting symmetries in the ground states of
${t_1}$-${t_2}$-${J_1}$-${J_2}$ model on square lattice. The
superconducting symmetries of the ground states are determined by
the frustration amplitude $t_2/t_1$ and doping concentration. The
phase diagram of this system in frustration-doping plane is given.
The order of the phase transitions among these different
superconducting symmetry states of the system is discussed.

\end{abstract}

\pacs{74.20.Rp, 74.25.Dw, 74.20.Mn}

\maketitle

Frustrated magnets based on transition metals have attracted much
theoretical and experimental effort in the past years, a variety
of exotic quantum effects\cite{Y.Taguchi,co,kos,huang} expected
when competitive interactions lead a system into a frustrated
state, where it is impossible to satisfy all the pair interactions
simultaneously. One of the most simple yet very important systems
is spin-1/2 Heisenberg model on a square lattice with competing
nearest $J_1$ and next-nearest $J_2$ neighbor spin interactions.
Despite its simplicity, it is not only in its own right
academically interesting, but also practically provides a window
for looking inside into the superconductivity mechanism. Recently
found vanadium-based compounds have provided experimental
realizations of $J_1$-$J_2$ model with $J_2/J_1\sim 1$
\cite{melzi,carretta,rosner,bombardi} and again have stimulated
research on this charming system.

As is known, a single layer in undoped cuprate high-$T_c$
compounds are well described by a square-lattice Heisenberg model
displaying long-range antiferromagnetic N\'{e}el order. The common
belief is that upon doping, this long range order is destroyed and
a different non-magnetic phase sets in, which, accompanied by
fluctuations, turns the system into superconducting phase. One
idea suggests that the effect of doping in destroying the N\'eel
order might be accounted for by the introduction of frustration in
original Heisenberg model \cite{inui}. Following this suggestion,
many works had been focused on finding such phases in frustrated
quantum magnets
\cite{chandra,dagotto,read-sachdev,chandra-coleman-larkin}.
Recently, the half-filled Hubbard model with both nearest neighbor
and next-nearest-neighbor hopping term has been investigated
\cite{nevidomskyy}. This model in the large $U$ limit is
equivalent to $J_1$-$J_2$ model. In that paper, variational
cluster approach shows that $d_{x^2-y^2}$-wave superconductivity
can also occur at half-filling when the Hubbard system is under
pressure provided that the frustration and the on-site repulsion
are not too large.

However, there is another issue which is worth while to be
investigated: possible superconducting symmetries of $J_1$-$J_2$
model under doping.

At half-filling $J_1$-$J_2$ model respectively exhibits two
long-range magnetic orders in two distinct limit: a) the N\'{e}el
phase in the limit of $J_2/J_1 \rightarrow 0$; b) the so called
collinear phase \cite{misguich} with ordering wave vector
$(\pi,0)$ or $(0,\pi)$ in the limit of $J_2/J_1 >> 1$. Frustration
makes the system highly degenerated in the intermediate range and
may dramatically suppresses the magnetic long range order
\cite{xu-ting,oguchi,igarashi,gochev,dotsenko}. Hence together
with doping, frustration may provide the system exotic
superconducting symmetries over a broad range of doping and
frustration.

In this work we investigate superconducting symmetry of the ground
state of doped ${J_1}$-${J_2}$ model, i.e. $t_1$-$t_2$-$J_1$-$J_2$
model. To investigate the properties of the ground state, a simple
yet powerful method is the renormalized mean-field theory (RMFT)
\cite{anderson-lee,zhang88} in which the kinetic and superexchange
energies are renormalized by different doping-dependent factors
$g_t$ and $g_s$, respectively. Despite of the simplicity of this
method, it can lead to semi-quantitative even quantitative
explanation of some ground state properties of cuprate
superconductors \cite{anderson-lee,zhang-rice,zhang88}. In this
letter, with the help of RMFT, we show that the superconducting
symmetry of the ground state varies among different types when
tuning the frustration amplitude and doping concentration of the
system.

{\it Model} --- The Hamiltonian of the $t_1$-$t_2$-$J_1$-$J_2$
model takes the form of
\begin{eqnarray}
H&=&\hat{P_d}H_t\hat{P_d}+H_s\;\;, \nonumber\\
H_t& =& -t_1\sum_{\langle
nn\rangle\sigma}c^{\dag}_{i\sigma}c_{j\sigma}
-t_2\sum_{\langle nnn\rangle\sigma}c^{\dag}_{i\sigma}c_{j\sigma}+h.c.\;\;, \nonumber\\
H_s& =& J_1\sum_{\langle nn\rangle}\mathbf{S}_{i}\cdot
\mathbf{S}_{j}+J_2 \sum_{\langle nnn\rangle}\mathbf{S}_{i}\cdot
\mathbf{S}_{j}\;\;,
\end{eqnarray}
$\hat{P_d}=\prod\limits_{i}(1- n_{i\uparrow }n_{i\downarrow })$ is
the Gutzwiller projection operator \cite{zhang88} which removes
totally the doubly occupied states. $t_1$ and $t_2$ are
nearest-neighbor and next-nearest-neighbor hopping amplitude. When
they are all positive, the Hamiltonian represents hole doping
case. The electron doping case can be achieved via particle-hole
transformation, changing the sign of $t_2$ while keeping the sign
of $t_1$ unchanged \cite{liu-trivedi}. $J_1$ and $J_2$ are
respectively the nearest-neighbor and next-nearest-neighbor
antiferromagnetic coupling constants, they raise frustration in
the system. We use $t_1$ as the energy unit and set
$\mathrm{t}_1/\mathrm{J}_1=3$ for conventional reason
\cite{kim-shen,coldea}, ${J_2}/{J_1}=({t}_2/{t}_1)^{2}$ since the
superexchanges have relations of ${J}= 4{t}^2/U$ with hopping
parameters in the large Hubbard ${U}$ limit. And  we take
$\eta={t_2}/{t_1}$ as frustration amplitude.

{\it Method} --- Renormalized mean-field theory. In the frame of
RMFT, to investigate the ground state of the above mentioned
Hamiltonian, the trial state is suggested to be a projected $BCS$
state $|\Psi\rangle=P_d|\Psi_{0}\rangle$, where $|\Psi_{0}\rangle
=
\prod_{k}(u_k+\upsilon_{k}c^\dagger_{k\uparrow}c^\dagger_{-k\downarrow})|0\rangle$.
And the projection operator is taken into account by a set of
renormalized factors \cite{gutzwiller,vollhardt,ogawa}, i.e. we
have $\langle\psi|H|\psi\rangle=\langle\psi_0|H'|\psi_0\rangle=
\langle\psi_0|g_tH_t+g_sH_s|\psi_0\rangle$. In homogenous case the
renormalized factors \cite{zhang88} $g_{t}=2\delta /(1+\delta)$
and $g_{s}=4/(1+\delta)^2$.

\begin{figure}
\includegraphics[width=0.88\columnwidth]{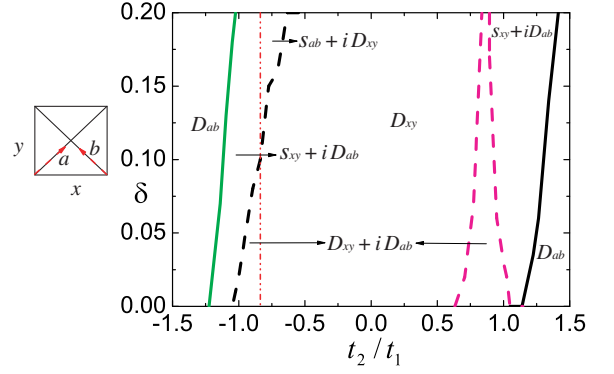}
\caption{Left picture is schematic structure of
${t_1}$-${t_2}$-${J_1}$-${J_2}$ model lattice. We use $a,b$ to
distinguish the two different orientation diagonal bond. Right
picture is phase diagram of this lattice, dashed lines are
boundary of first order phase transition.}\label{1}
\end{figure}

Minimize the quantity $W=\langle H'-\mu
\sum_{i\sigma}c_{i\sigma}^{\dagger}c_{i\sigma}\rangle_0$ with
respect to $u_k$ and $\upsilon_k$, and introduce two mean-field
parameters $\Delta_\tau=\langle
c^{\dag}_{i\uparrow}c^{\dag}_{i+\tau\downarrow}-c^{\dag}_{i\downarrow}c^{\dag}_{i+\tau\uparrow}\rangle_0
$ and $\xi_\tau=\sum_{\sigma}\langle
c^{\dag}_{i\sigma}c_{i+\tau,\sigma}\rangle_{0}$, where $\tau$
indicates four different bond directions sketched in Fig.\ref{1},
we get the coupled gap equations as follows
\begin{eqnarray}\label{gap equ2}
 \Delta_\tau&=&N_{s}^{-1}\sum_{k}\cos{k_\tau}\Delta_{\vec{k}}/E_{\vec{k}}\;\;,\\
 \xi_\tau&=-&N_{s}^{-1}\sum_{k}\cos{k_\tau}\xi_{\vec{k}}/E_{\vec{k}}\;\;,
\end{eqnarray}
where
$\Delta_{\vec{k}}=\sum_{\tau}\Delta_{\tau}{J}_{\tau}/{J}_1\cos{k_{\tau}}$,
$\xi_{\vec{k}}=\bar{\varepsilon}_{\vec{k}}-\sum_{\tau}\xi_{\tau}{J}_{\tau}/{J}_1\cos{k_{\tau}}$,
$E_{\vec{k}}=\sqrt{\xi_{\vec{k}}^2+|\Delta_{\vec{k}}|^2}$, $N_s$
is the total number of sites. In the above equations $
\bar{\varepsilon}_{\vec{k}}=(-2g_{t}{t}_1\sum_{\tau}{t}_{\tau}/{t}_1
\cos{k_{\tau}}-\tilde{\mu})/(\frac{3}{4}g_{s}{J})$ and
$\tilde{\mu}=\mu+N_{s} ^{-1}\langle \frac{\partial H'}{\partial
\delta}\rangle_0. $ These gap equations should be solved
simultaneously with
$\delta=N_{s}^{-1}\sum_{\vec{k}}\xi_{\vec{k}}/E_{\vec{k}} $, the
resulting $\Delta$'s determine the symmetry of possible
superconductivity.

Although the RMFT cannot provide us a true picture of the system
in exact half-filling case, the symmetries of gap parameters
obtained at that point will not change under small doping
\cite{anderson-lee}. Thus, in order to investigate possible
superconducting symmetries of the system under doping, our
strategy is following: at first, we solve the gap equation at half
filling, find out the possible symmetries of the mean-field
parameters. These symmetry states may degenerate at half-filling.
Then we switch on the doping, compare the energies of different
symmetry states, we can find the true ground states for different
doping level and frustration amplitude.

At half filling, energy if per site is
$-\frac{3}{8}g_{s}J\sum_{k}E_k$, with the ansatz that $E_k$ of
ground state energy can be written as a summation of square terms
of cosine functions, after some calculation, we find that several
symmetry states come out, including: (1) $D_{xy}+is_{ab}$-wave
state degenerates with $D_{xy}+iD_{ab}$-wave state, both of them
have $\xi_{a,b}=0$; (2) $D_{ab}+iD_{xy}$-wave state degenerates
with $D_{ab}+is_{xy}$-wave state, $\xi_{x,y}=0$ in both states.
Here $i$ means that the difference between the phase $\phi_a$ of
$\Delta_{a}$ and the phase $\phi_x$ of $\Delta_{x}$ is
$|\phi_{a}-\phi_{x}|=\pi/2$, we use $D_{\tau_1\tau_2}$ and
$s_{\tau_1\tau_2}$ to denote $d$-wave symmetry and $s$-wave
symmetry on direction $\tau_1$ and $\tau_2$ respectively.

When changing the doping level and the frustration amplitude,
these superconducting symmetries compete, and energetically, each
occupies a specific region in the phase diagram, as shown in
Fig.\ref{1}. In this phase diagram dashed line are boundary of the
first order phase transition where mean-field parameters change
suddenly. The bold lines show second order phase transitions. Left
picture in Fig.\ref{1} is a cartoon sketch of the lattice
structure that we discuss, $a$ and $b$ denote the two different
diagonal bond. In the following, we are going to discuss the phase
diagram for positive and the negative $t_2/t_1$ case in more
detail.

\begin{figure}
\includegraphics[width=0.88\columnwidth]{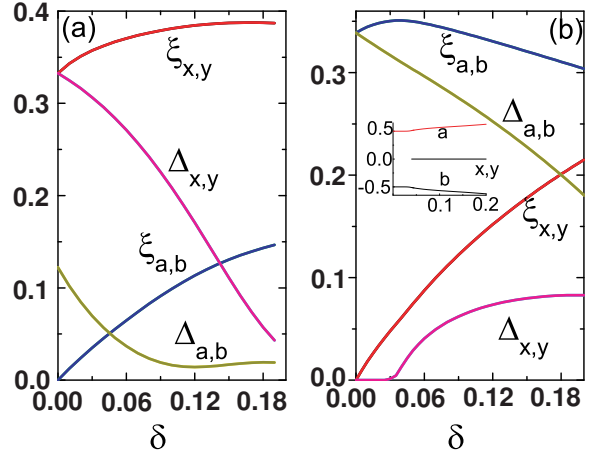}
\caption{Figure(a) shows mean-field parameters of $D_{xy}+iD_{ab}$
state as function of $\delta$ for $\eta=\sqrt{0.6}$. Figure(b)
shows parameter functions for $\eta=\sqrt{1.5}$, the inset picture
shows phase of pairing parameters, its unit is $\pi$. }\label{2}
\end{figure}

{\it Positive $\eta$  cases. }--- For $\eta<\sqrt{0.5}$, as
expected, although the next nearest neighbor interaction
frustrates the system, it does not destroy the $D$-wave
superconducting symmetry state  with $\Delta_x=-\Delta_y$ and
$\Delta_{a,b}=0$ for all doping level under investigation. For
$\eta>\sqrt{0.5}$, pairing on diagonal links should be considered
seriously.

RMFT calculation shows that when the frustration amplitude falls
in the range of $\sqrt{0.5}\leq\eta<\sqrt{0.7}$,
$D_{xy}+iD_{ab}$-wave state with $|\Delta_{x,y}|>|\Delta_{a,b}|$
is the most energetically favored state in the small doping
region. In order to show the dependence on the doping
concentration, as an example, we take $\eta=\sqrt{0.6}$ and plot
$D_{xy}+iD_{ab}$-wave parameters in Fig.\ref{2}(a). With
increasing $\delta$, $\Delta_{a,b}$ decreases rapidly. One can see
from phase diagram Fig.\ref{1} that after $\delta=0.06$,
$D_{xy}$-wave state is energetically more favored, and there is a
sudden change of $\Delta_{a,b}$ in this transition.

\begin{figure}
\includegraphics[width=0.88\columnwidth]{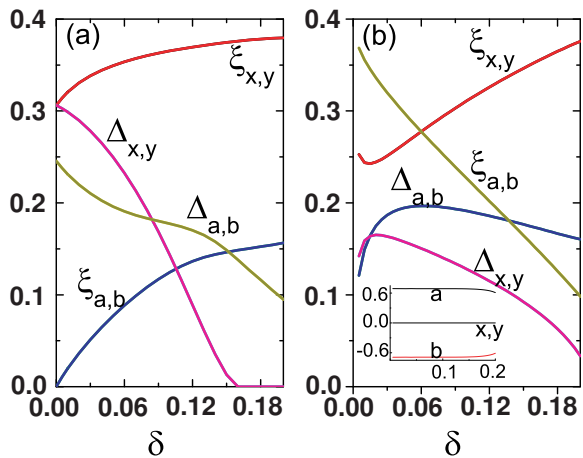}
\caption{(color online) For $\eta=\sqrt{0.9}$, Fig.(a) shows
mean-field parameters of $D_{xy}+iD_{ab}$-wave state, while
Fig.(b) shows that of $s_{xy}+iD_{ab}$-wave, the inset of (b)
shows the phases of its pairing parameters.}\label{3}
\end{figure}

For about $\sqrt{0.8}<\eta<\sqrt{1.1}$ in small doping level
stable state is still $D_{xy}+iD_{ab}$-wave state, however, in
relatively high doping level another mixed superconducting
symmetry state of $s_{xy}+iD_{ab}$-wave dominates. Again, in order
to illustrate the dependence on doping level, we take
$\eta=\sqrt{0.9}$ and plot the parameters for $D_{xy}+iD_{ab}$
symmetry and $s_{xy}+iD_{ab}$ symmetry in Fig.\ref{3}(a) and
Fig.\ref{3}(b), respectively. Inset of Fig.\ref{3}(b) shows the
phases of $\Delta_{\tau}$ in different bond directions, and
clearly reflects the symmetry property of the
$s_{xy}+iD_{ab}$-wave state. From this inset, we can also see that
the phases of $\Delta_{a,b}$ is slightly away from $\pi/2$. By
comparing the energies of these two different states, the
calculation shows that a first order phase transition from
$D_{xy}+iD_{ab}$-wave state to $s_{xy}+iD_{ab}$-wave state occurs
when increasing the doping level $\delta$ across $0.06$. When
$\eta$ is tinily larger than $1$, in very small doping region
stable state most likely is another type of $D_{ab}+iD_{xy}$-wave
with $|\Delta_{a,b}|>|\Delta_{x,y}|$, it occupies really an
extremely small area and can not be shown explicitly in our phase
diagram.

In the region of $\sqrt{1.2}<\eta<\sqrt{2}$, stable state is
$D_{ab}$-wave in small doping and will change to
$s_{xy}+iD_{ab}$-wave state when increasing the doping level with
the corresponding phase transition being second order. In
Fig.\ref{2}(b) we show the case of $\eta=\sqrt{1.5}$, it is clear
that the parameters vary smoothly from $D_{ab}$-wave type to
$s_{xy}+iD_{ab}$-wave type when increasing $\delta$. Inset of
Fig.\ref{2}(b) clearly shows that in $s_{xy}+iD_{ab}$-wave state
phase difference of $\Delta_{a,b}$ is not exactly $\pi$ and varies
slowly with doping concentration.

For very large $\eta$, it is reasonable to expect a state with
$d$-wave pairing only on diagonal bonds, i.e. the state is
$D_{ab}$-wave.

{\it Negative $\eta$ case. }---  As in the positive $\eta$ case,
when $|\eta|$ takes small value, roughly smaller than
$\sqrt{0.5}$, superconducting symmetry of the system is
conventional $d$-wave on plaquette bonds.

\begin{figure}
\includegraphics[width=0.88\columnwidth]{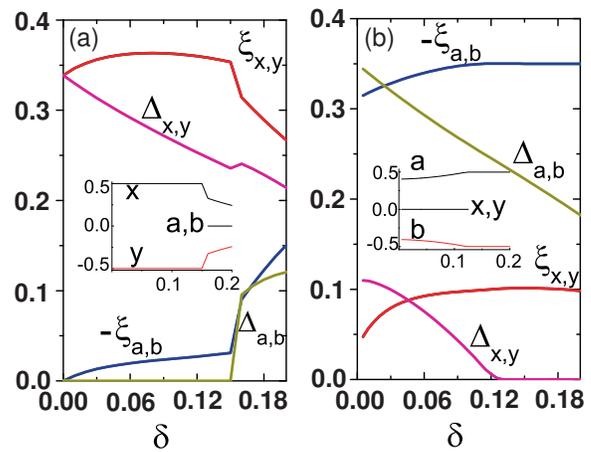}
\caption{(color online) Figure(a) shows mean-field parameters
amplitude of stable state as function of $\delta$ for
$\eta=-\sqrt{0.6}$, figure(b) shows parameter function of stable
state for $\eta=-\sqrt{1.2}$, the inset of (a,b) shows the phase
of $\Delta$.}\label{f1.26}
\end{figure}

For $-\sqrt{0.7}<\eta<-\sqrt{0.5}$, the $D_{xy}$-wave symmetry
changes into $s_{ab}+i D_{xy}$-wave state when increasing doping
level with the phase transition being first order. It should be
emphasized that the phase difference between $\Delta_x$ and
$\Delta_a$ in the latter state is less than $\pi/2$ and decreases
with increasing doping level. Fig.\ref{f1.26}(a) shows that for
$\eta=-\sqrt{0.6}$, when $\delta>0.16$, $D_{xy}$-wave state
discontinuously changes to $s_{ab}+i D_{xy}$-wave state. Inset
shows the phase of pairing parameters, at the critical point
symmetry of superconductivity changes suddenly.

When $-\sqrt{1.5}<\eta<-\sqrt{1.1}$, $D_{ab}$-wave is the stable
state in high doping level while in small doping level stable
state is $s_{xy}+i D_{ab}$-wave symmetry. $\eta=-\sqrt{1.2}$ case
is shown explicitly in Fig.\ref{f1.26}(b) as an example. With
increasing $\delta$ pairing on diagonal bonds $\Delta_{x,y}$
decrease to zero rapidly and $s_{xy}+i D_{ab}$-wave state varies
to $D_{ab}$-wave smoothly through a second order phase transition.
$D_{ab}$-wave state occupies more range of $\delta$ when
increasing $|\eta|$ till $|\eta|=\sqrt{1.5}$, after that $D_{ab}$
is the only stable one for all doping level.

\begin{figure}
\includegraphics[width=0.88\columnwidth]{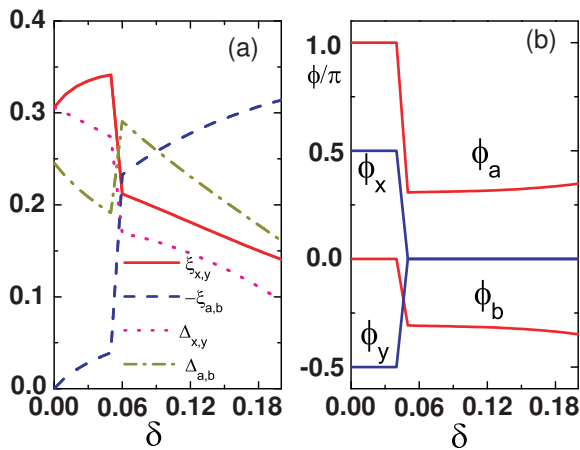}
\caption{(color online) Figure(a) shows mean-field parameters
amplitude of stable state as function of $\delta$ for
$\eta=-\sqrt{0.9}$, figure(b) shows phase of pairing parameters.
At the critical point symmetry of superconductivity change for
$D_{xy}+iD_{ab}$ to $s_{xy}+iD_{ab}$ wave state. }\label{5}
\end{figure}

For $-\sqrt{1.1}<\eta<-\sqrt{0.7}$, in small doping level
$D_{xy}+iD_{ab}$ state has more energy gain while $s_{xy}+iD_{ab}$
is more stable at higher doping level. Fig.\ref{5}(a) shows how
parameter amplitudes varied with doping concentration at
$\eta=-\sqrt{0.9}$ for best energy stable state
 and Fig.\ref{5}(b) shows the phase changing of stable state. The
 sudden changes of the parameter amplitudes and phases clearly
 indicate that the phase transition here is first order.

{\it Summary and discussion} ---In summary, we have investigated
the  possible superconducting symmetries of frustrated
${t_1}$-${t_2}$-${J_1}$-${J_2}$ model by RMFT method. In terms of
frustration amplitude $t_2/t_1$ and doping concentration $\delta$
a phase diagram for possible superconducting symmetry states has
been presented. We have shown that in weakly and strongly
frustrated cases pure $d$-wave states are on plaquette or diagonal
bonds, respectively. When $t_1$ and $t_2$ are comparable, pairing
parameters on plaquette bonds and diagonal bonds are both finite
and mixed symmetry states such as $D+iD$, $D+is$ appear. Each
symmetry state occupies a specific region in the phase diagram,
transitions between these states are the first order or second
order, the latter one corresponding to relatively large $J_2$. The
two kinds of time-reversal broken mixed states are consistent with
those found in two-dimensional Fermi liquid with attractive
interaction in both $s$ and $D$ channel~\cite{fermi}, $D+is$
time-reversal broken state may also appear in the pseudogap
phase~\cite{timere}. We hope that our calculation may shed some
new light on the study of mechanism of superconducting order in
high-$T_c$.

{\it Acknowledgment} --- Work supported by NSF of china
Nos.10747145, and 10575068.


\begin{references}
\bibitem{Y.Taguchi}Y. Taguchi, Y. Oohara, H. Yoshizawa, N. Nagaosa, and
Y. Tokura, Science {\bf 291}, 2573 (2001).
\bibitem{co} K. Takada, H. Sakurai, E. Takayama-Muromachi, F. Lzumi, R. A. Dilanina, and T. Sasaki, Narure(London)
{\bf 422}, 53 (2003).

\bibitem{kos} S. Yonezawa, Y. Muraoka, Y. Matsushita, and Z. Hiroi, J. Phus :condens. Matter, {\bf 75}, L9
(2004).
\bibitem{huang} H. X. Huang, Y. Q. Li, J. Y. Gan, Y. Chen and F. C. Zhang, Phys. Rev. B {\bf 75}, 184523
(2007).

\bibitem{melzi} R. Melzi, P. Carretta, A. Lascialfari, M. Mambrini, M. Troyer,
P. Millet and F. Mila, Phys. Rev. Lett. {\bf 85}, 1318 (2000).

\bibitem{carretta} P. Carretta, N. Papinutto, C. B. Azzoni, M. C. Mozzati,
E. Pavarini, S. Gonthier, and P. Millet, Phys. Rev. B {\bf 66},
094420 (2002); P. Carretta, R. Melzi, N. Papinutto and P. Millet,
Phys. Rev. Lett. {\bf 88}, 047601 (2002).

\bibitem{rosner}H. Rosner, R. R. P. Singh, W. H. Zheng, J. Oitmaa, S.-L. Drechsler,
and W. E. Pickett, Phys. Rev. Lett. {\bf 88}, 186405 (2002); H.
Rosner, R. R. P. Singh, W. H. Zheng, J. Oitmaa, and W. E. Pickett,
Phys. Rev. B {\bf 67}, 014416 (2003).


\bibitem{bombardi} A. Bombardi, J. Rodriguez-Carvajal, S. Di Matteo, F. de Bergevin,
L. Paolasini, P. Carretta, P. Millet, and R. Caciuffo, Phys. Rev.
Lett. {\bf 93}, 027202(2004).


\bibitem{inui} M. Inui, S. Doniach, and M. Gabay, Phys. Rev. B
{\bf 38}, 6631 (1988).

\bibitem{chandra} P. Chandra and B. Doucot, Phys. Rev. B {\bf 38},
9335 (1988).

\bibitem{dagotto} E. Dagotto and A. Moreo, Phys. Rev. Lett. {\bf
63}, 2148 (1989).

\bibitem{read-sachdev} N. Read and S. Sachdev, Phys. Rev. Lett.
{\bf 62}, 1694 (1989).

\bibitem{chandra-coleman-larkin} P. Chandra, P. Coleman, and A.I.
Larkin,  Phys. Rev. Lett. {\bf 64}, 88 (1990).

\bibitem{nevidomskyy} A. H. Nevidomskyy, C. Scheiber, D. S\'{e}n\'{e}chal,
and A.-M. S. Tremblay  Phys. Rev. B {\bf 77}, 064427 (2008).


\bibitem{misguich} G. Misguich and C. Lhuillier, arXiv:
cond-mat/0310405 (2003).


\bibitem{xu-ting} J. H. Xu and C. S. Ting, Phys. Rev. B {\bf 42}, 6861 (1990).

\bibitem{oguchi} T. Oguchi and H. Kitatani, J. Phys. Soc. Jpn.  {\bf 59}, 3322 (1990).

\bibitem{igarashi} J. Igarashi, J. Phys. Soc. Jpn.  {\bf 62}, 4449 (1993).

\bibitem{gochev} I. G. Gochev,  Phys. Rev. B {\bf 49}, 9594 (1994).
\bibitem{dotsenko} A. V. Dotsenko and O. P. Sushkov,  Phys. Rev. B {\bf 50}, 13821 (1994).


\bibitem{anderson-lee} P. W. Anderson, P. A Lee, M. Randeria, T. M. Rice,
N. Trivedi and F. C. Zhang, J. Phys. Cond. Matt. {\bf 16}, R755
(2004).

\bibitem{zhang88}  F. C. Zhang, C. Gros, T. M. Rice and H. Shiba,  Supercond. Sci.
Tech. \textbf{1}, 36 (1988).


\bibitem{zhang-rice} F. C. Zhang and T. M. Rice, Phys. Rev. B {\bf
37}, 3759 (1988).
\bibitem{liu-trivedi} J. Liu, N. Trivedi, Y. Lee, B. N. Harmon, and
J. Schmalian, Phys. Rev. Lett. {\bf 99}, 227003 (2007).


\bibitem{kim-shen} C. Kim, P. J. White, Z.-X. Shen, T. Tohyama, Y.
Shibata, S. Maekawa, B. O. Wells, Y. J. Kim, R. J. Birgeneau, and
M. A. Kastner, Phys. Rev. Lett. {\bf 80}, 4245 (1998).

\bibitem{coldea} R. Coldea, S. M. Hayden, G. Aeppli, T. G. Perring,
C. D. Frost, T. E. Mason, S. -W. Cheong, and Z. Fisk, Phys. Rev.
Lett. {\bf 86}, 5377 (2001).

\bibitem{gutzwiller} M. C. Gutzwiller, Phys. Rev. {\bf 137}, A1726(1965).

\bibitem{vollhardt} D. Vollhardt, Rev. Mod. Phys. {\bf 56}, 99 (1984).

\bibitem{ogawa} T. Ogawa, K. Kanda and T. Matsubara, Prog. Theor.
Phys. {\bf 53}, 614 (1975).


\bibitem{aoki} H. Aoki, J. Phys: Condensed Matter {\bf 16}, V1 (2004).

\bibitem{fermi} K. A. Musaelian, J. Betouras, A. V. Chubukov and
R. Joynt, Phys. Rev. B. {\bf 53}, 3598 (1996).

\bibitem{timere} R. P. Kaur and D. F. Agterberg, Phys. Rev. B {\bf 68}, 100506(R) (2003).

\end{references}
\end{document}